\documentstyle[11pt,aaspp4]{article}
\received{21 April 1997}

\input psfig.sty

\def\simgt{\lower.5ex\hbox{$\; \buildrel > \over \sim \;$}}
\def\simlt{\lower.5ex\hbox{$\; \buildrel < \over \sim \;$}}
\def\sp{\hspace{1.5pt}}

\def\src{{1E\sp1841$-$045}}
  
\lefthead{Gotthelf \& Vasisht}
\righthead{Supernova Remnant Kes\sp73}

\begin{document}

\title{Kes\sp73: A Young Supernova Remnant with an X-ray Bright, Radio-quiet Central Source}

\author{E. V. Gotthelf\altaffilmark{1}}
\affil{NASA/Goddard Space Flight Center, Greenbelt, MD 20771}

\author{G. Vasisht}
\affil{California Institute of Technology, MS 105-24, Pasadena, CA 91125}

\altaffiltext{1}{Also: Universities Space Research Association}

\begin{abstract}

We clarify the nature of the small-diameter supernova remnant (SNR)
Kes\sp73 and its central compact source, 1E\sp1841$-$045, using X-ray
data acquired with the ASCA Observatory.  We introduce a
spatio-spectral decomposition technique necessary to disentangle the
ASCA spectrum of the compact source from the barely resolved
shell-type remnant. The source spectrum (1 -- 8 keV) is characterized
by an absorbed power-law with a photon index $\alpha \simeq 3.4$ and
$N_H \simeq 3.0\times 10^{22}$ cm$^{-2}$, possibly non-thermal in
nature. This bright X-ray source is likely a slowly spinning pulsar,
whose detection is reported in our companion paper (Vasisht \&
Gotthelf 1997). The SNR spectrum is characteristic of a thermal
plasma, with $\rm{kT} \simeq 0.7$ keV, and emission lines typical of a
young remnants. The element Mg and possibly O and Ne are found to be
over-abundant, qualitatively suggesting an origin from a massive
progenitor. We find that Kes\sp73 is a young ($\simlt 2000$ yr) type
II/Ib SNR containing a neutron star pulsar spinning anomalously slow
for its age. Kes\sp73 is yet another member of a growing class of SNRs
containing radio-quiet compact sources with a hard spectral signature.

\end{abstract}

\keywords{stars: individual (Kes\sp73, 1E\sp1841$-$045) --- stars: neutron --- 
supernova remnants --- X-rays: stars}

\section{Introduction}

The supernova remnant Kes\sp73 is a limb - brightened, shell-type radio
remnant $\sim 4'$ in diameter, located along the Galactic plane
(G\sp27.4+0.0). The X-ray remnant is comparable in size to the bright
radio shell, but is dominated by localized emission knots, interpreted
as clumpy, diffuse emission in the interior, possibly due to
fluorescence from reverse shock (Helfand et al. 1994 using ROSAT HRI
images). Centered on the remnant is the unresolved compact X-ray
source, 1E\sp1841$-$045, discovered with the Einstein Observatory
(Kriss et al. 1985). Like the supernova remnants RCW 103 and CTB 109, 
the radio observations show no evidence for a plerionic component or a
localized radio counterpart to \src, with a flux limit $F_{6cm} < 0.6$
mJy (Kriss et al.  1985). An HI absorption-based distance
determination shows that Kes\sp73 lies $\sim 7.0$ kpc away
(Sanbonmatsu \& Helfand 1992).

The nature of Kes\sp73 and in particular, its compact source, has been
considered extensively in two earlier studies.  Kriss et al. (1985)
were unable to distinguish between thermal and non-thermal emission
from \src; they suggested as the origin of the emission either a hot
thermal neutron star (NS), a crab-like pulsar, a plerionic nebula, or
an accretion powered binary.  A more recent study by Helfand et al.
(1994) reconsidered these possibilities using deep ROSAT HRI
pointings. They inferred a hard spectral component for \src, placed a
limit on its long term variability of a factor of two change in flux,
and suggested an accreting binary origin for the central source.

In this {\it Letter}, we use the broad-band ASCA X-ray observations of
Kes\sp73 to further clarify its nature. We present a novel technique
for isolating the background subtracted spectrum of the central source. We
show that the spectrum of \src\ is represented by a steep power law,
possibly of non-thermal origin, and suggest that  Kes\sp73 is a young
($\simlt 2000$ yrs) Type II/Ib SNR, the product of a massive progenitor.
In our companion paper (Vasisht \& Gotthelf, referred herein as GV97)
we report our discovery of a 11.8 s periodicity from \src, likely from
an anomalous pulsar, the stellar remnant of the supernova explosion.
The date and typing of Kes 73 are critical to understand the nature of
the pulsar. We present Kes\sp73 as another member of an emerging class 
of young thermal remnants which are found to contain a radio-quiet, 
hard X-ray point source.

\section{Observations}

ASCA (Tanaka et al. 1994) conducted a two day observation of Kes 73 on
11-12 Oct 1993. In this study, we use data acquired with the two
Solid State Imaging Spectrometers (SIS-0 and SIS-1), made available from
the public archive. These sensors are sensitive to X-rays in the
$0.4-10.0$ keV band, with a nominal spectral resolution of 2\% at 6
keV ($\sim E^{-1/2}$).  The spatial resolution is limited by the X-ray
mirrors, whose azimuthally averaged point spread function (PSF) is
characterized by a narrow core of FWHM $50''$, and extended wings that
result in a half-power diameter of $3'$ (Jalota et al. 1993).  The
SIS data were acquired in 1-CCD mode, read out every 4 sec, using a
combination of FAINT and BRIGHT data modes. The target was centered on
the $11'\times 11'$ field-of-view of SIS-0 CCD-1.  All data were edited
using the standard REV1 screening criteria which resulted in an effective
exposure of $\sim 34$ ksecs per sensor. A description of the data acquired
with the other two focal plane instruments is presented in our
companion paper (GV97).


The broad-band image of Kes 73 from the combined SIS cameras contains
93 kcounts, with a count rate of 1.38 cps.  In Plate 1
we display images of the Kes\sp73 region in the soft- and hard-band,
below and above 2.5 keV, respectively. The images are centered on the
point-like source, whose position is consistent with that of 1E\sp1841$-$045,
to within the formal SIS error circle (Gotthelf 1996). The hard-band image
clearly shows the clover-leaf shaped PSF, consistent with an
unresolved point source, whereas the soft-band image suggest an additional
diffuse structure.  The FWHM of the normalized brightness distribution
profiles in the two bands are $44^{\prime\prime}$ (hard) and
$108^{\prime\prime}$ (soft) (see Fig 1).  This difference is more
apparent in the deconvolved images in the lower panel of Plate 1.
These clearly show that the thermal X-ray shell is contained in the
soft-band and surrounds the harder, bright compact source.  The images
have been restored with 50 iterations of the Lucy algorithm (Lucy
1974) to deconvolve the complex shape of the PSF, and are displayed in
an identical fashion.

An image of the SNR nebulosity is made by subtracting the compact
component from the soft-band image. For this, we used as a model PSF
the SIS image of the moderately bright point-like source EX Hydra,
normalized by the derived source spectrum (see below) to estimate the
relative contribution of the compact source to the total emission
below 2.5 keV. The subtracted SIS image closely follows the brightness
distribution in the smoothed Rosat HRI image (Plate 1c), reproducing the
three areas of enhanced emission.  Other features are easily ascribed
to differences in the spectral response of the two instruments. See
Hwang \& Gotthelf (1997) for a discussion on interpreting features in
the processed images.


\section{Isolating the compact source spectrum}

The broad winged ASCA PSF does not allow spatial isolation of the
point-source emission from that of the SNR. Instead we adopt a
spatio-spectral decomposition method detailed in Wang \& Gotthelf
(1997) (referred herein as WG97) to separate the spectra of the two
components.  We then use the decomposed point-source spectrum to: (i)
verify and consolidate the fit to the point-source spectrum in the
total spectrum of the Kes\sp73 region (see \S 4.2) and (ii) compute
fluxes and luminosities of the point-source. Below we outline a
modified version of the WG97 method, germane to a point source
embedded within a SNR shell.

We start with the assumption that the spatial distribution of the
Kes\sp73 emission consists of two components, i.e., a point-source
embedded in a diffuse nebula. We then decompose the source spectrum
from the underlying nebula (+ background), by simultaneously solving
for the source and nebular spectra using the ratio of observed and
expected counts in concentric regions; we search for deviations from
the radial-average profile centered on the source, from that expected
for a point-source.

The original method, discussed in WG97, assumes a uniform
background. From the Rosat HRI morphology, we know this is not the
case for Kes\sp73, for which the background consists of both the
thermal-shell emission and the field-background.  Therefore, we use
the HRI data to estimate a correction factor to a uniform background
from the relative counts in the source and nebular region.  Our
procedure is as follows: (i) We first extract spectra from two annuli
centered on the source (see Fig. 1). The first annulus is a small
circle ($r = 1\farcm2$) encompassing the central pulsar counts. The
second annulus ($ 1\farcm5 > r \geq 2\farcm7$) is chosen to enclose
the bulk of the shell emission. (ii) Next, we compute the expected
radial profile for a point-source in these regions from a similar
1-CCD mode observation of EX Hydra.  To compensate for the non-uniform
SNR distribution, we compute a nebular correction factor using the
relative counts per pixel between the center and edge of the projected
HRI X-ray nebula (iii) With the above information, we decompose the
spectra in the two annuli into a source spectrum and a nebular
spectrum using eqns. A2 3-4 of WG97, applied to each spectral channel.

The spectra separated into two unmixed components: a line dominated
spectra expected from thermal emission of a shocked SNR shell and a
steep power-law continuum spectrum for the compact source. These are
shown on Plate 1a \& 1b. No coercion or prejudice is used to force the
clean separation into the distinct nebula and source spectra. The
background subtracted source spectrum (Plate 1a) accounts for the
harder ($> 2.5$ keV) emission and provides an independent absorption
measurement. The spectrum is trivially fitted with an absorbed
power-law with a photon index $\Gamma = 3.4 \pm 0.3$ and $N_H
\simeq 3.0 \pm 0.4\times 10^{22}$ cm$^{-2}$.  The inferred unabsorbed
luminosity is ${L_X} (1.0 - 10.0 \ \rm{keV}) \sim 3 \times
10^{35} \ d^2_7$ erg s$^{-1}$, for an assumed distance of $7.0 \ d_7$
kpc (see Table 1).  The spectral shape ($\propto E^{-2}$) is in
accordance with the steep power-law spectra seen in other anomalous
X-ray pulsars (see Corbet et al. 1995). We now use this fit to
constrain the power-law emission component in the combined
fit to the Kes\sp73 spectrum.

\section{The SNR nebula spectrum}

The SIS energy spectrum photons were selected from within a circular
emission region of radius $\simeq 4'$, limited by the size of the CCD.
The broad ASCA PSF makes background estimation extremely difficult,
since the source flux from a diffuse object extends over most of the
CCD chip. For this analysis, we extract a background spectrum from
nearby archival pointings of the Galactic ridge. The resulting
background-subtracted spectrum of the shell plus compact source is
shown in Fig 2. The line-dominated spectrum, typical of those seen
from SNRs, suffers high foreground absorption in the energy range below
$\sim 1.5$ keV.  In the energy range spanning $\sim 1 - 10$ keV, we see
K-shell emission from highly ionized ions of Mg, Si, S, Ca, and Ar. The
identified emission lines and their characteristic parameters are
displayed in Table 2.

Collisional ionization equilibrium (Raymond \& Smith 1977; Mewe et al.
1985 and references therein) and thermal bremsstrahlung models with
Gaussians for line emission, are fit to the spectra.  There are large
residuals, in either case, in the hard-band (2.5 - 10.0 keV)
suggesting an extra emission component. The image analysis, spanning
that energy range, clearly demonstrates that this component is mostly
emission from the compact source, for which we add a single power-law
to the fit.  A bremsstrahlung continuum, with Gaussians and a
power-law tail provides the best fit (see Table 1). In contrast, the
single temperature Raymond-Smith model with fixed relative metal
abundances and a power-law, is a relatively poor fit; it shows large
negative residuals mainly near the over-abundant Mg feature, and a
relatively large metal abundance of $\simeq 2.0$ cosmic, driven mainly
by the strong Si (1.83 keV) feature. In addition, we note excess flux 
in all our fits, at the lower energy range, $0.5 - 0.9$ keV, which we 
tentatively ascribe to O and Ne.

\section{Discussion}

A lower limit on the SNR age can be derived assuming free expansion of
the spherical remnant. For a typical maximum velocity of $10^4$ km
s$^{-1}$ seen in Type II SN, the projected size of Kes\sp73, ${R}_s 
\simeq 4.7 \ {d}_7$ pc, contrains its age to be $ \tau_s < 470$ yrs.
Since most of the thermal emission is due to shocked matter of
electron temperature ${kT_e} = 0.8$~keV, we compute a shock speed $v_s
= (16\rm{kT_e}/3\mu m_p)^{1/2}$ $\sim 900$ km s$^{-1}$.  This assumes
ion and electron equilibrium behind the shock ($\mu$ = 0.6 for cosmic
abundances). Inefficient electron heating would result in a larger
shock velocity, i.e., $v_s \simgt$ 900 km s$^{-1}$. If we assume that
the remnant has entered a well developed Sedov phase, ${R}_s =
2.5v_s\tau_s$, and therefore $\tau_s \simlt 2.2\times 10^3$ yr. This
age is further reduced if the remnant is not fully in Sedov
phase. From spectral fitting, the total thermal ($1 - 10$ keV)
luminosity of the SNR is ${L_X} \sim 3 \times 10^{35} d_7^2$ erg
s$^{-1}$ (this estimate excludes the uncertain, highly-absorbed
emission from O and Ne).  The instantaneous power radiated from
a shell is ${L_s(t)} = (16\pi/3) {R}_s^3(t){n}_o^2\Lambda (T)$, where
${n}_o$ is the mean pre-shock particle density and $\Lambda({T}) =
1.0\times 10^{-22} {T}_6^{-0.7} + 2.3\times 10^{-24}{T}_6^{0.5}$ erg
cm$^3$ s$^{-1}$ is the cooling function (McCray 1987), and ${T}_6$ is
the $\rm{kT_e}$ in units of 10$^6$ K.  We derive an ${n}_o$ $\sim$
0.8${d}_7^{-0.5}$ cm$^{-3}$. Under the strong-shock assumption the
post-shock electron density is $\sim 3$ cm$^{-3}$.

Kes 73 is evidently young, with the shocked plasma still ionizing and
thus non-equilibrium effects in the ionization balance are important.
We determined the diagnostic parameters for the non-equilibrium
ionization (NEI) plasma from the line intensity ratios of
He-like-K$\alpha$ to H-like-K$\alpha$, as well as, He-like-K$\beta$ to
He-like-K$\alpha$ ionic transitions of individual ions, Si and S in
this case.  The former ratio is a measure of the ionization degree and
is dependent upon both the electron temperature $\rm{kT_e}$ and the
ionization age $n_e$t, the product of the electron density and time
since the gas was last heated by the shock (Itoh 1977).  The latter
ratio, however, is solely a function of $kT_e$. The obtained
He-like-K$\beta$ to He-like-K$\alpha$ ratio for Si and S is $0.087 \pm
0.015$ and 0.089$\pm 0.039$, respectively. The He-like-K$\alpha$ to
H-like-K$\alpha$ intensity ratio of Si and S are $22 \pm 9$ and $15
\pm 9$, respectively.  We measure a $kT_e \simeq 0.75 - 0.90$ keV and
$0.71 - 0.97$ keV for Si and S, respectively, assuming a single
uniform plasma continuum and a power-law. These are in accordance with
the $\rm{kT_e}$ derived from the single temperature bremsstrahlung
continuum.  The ionization parameter was estimated to lie in the range
$n_e t \simeq 11.0 - 11.4$ (Masai 1984).  Using the post-shock density
inferred via normalization to the continuum component, we get $t =
\tau_s \sim 1800$ yr.  Several independent arguments, therefore,
suggest that Kes\sp73 is a young SNR, $\sim 2000$ yr-old, as indicated
by earlier studies (Helfand et al. 1994).

We estimate the total mass of the swept-up interstellar gas to be $M_s
\simeq 8.8 d_7^3$ M$_\odot$. This corresponds to typical envelop
masses ejected in a Type II supernova (progenitor mass $M > 8$
M$_\odot$), suggesting that the SNR dynamics could well still be in a
transitional stage between the free expansion and Sedov phases.  This
notion is qualitatively consistent with the development of a strong
reverse shock that can heat the metal-rich gas and cause the diffuse
emission to be seen in the SNR interior.  Additionally, the detection
of strong Mg, Si, S, and Ar emission and perhaps accompanying O and Ne
emission suggests ejecta-dominated gas. During their evolution,
massive stars are expected to produce large amounts of O-group
elements which are ejected during the supernova (Thieleman, Nomoto \&
Hashimoto 1994; Woosley 1991). These patterns have been observed in
the X-ray spectra of O-rich SNR (see Hughes \& Singh 1994; Hayashi et
al. 1994). Here the emission is highly absorbed and does not permit
quantitative estimates of O and Ne emission (although, upper-limits
and observed photon excess in the spectral range 0.4 - 1 keV are
consistent with abundant O and Ne).

The thermal spectrum of Kes\sp73 is remarkably similar to that of
two young, distant (therefore, high-absorbed) Type II/Ib SNRs,
G\sp11.2$-$0.3 and RCW\sp103. The former is a historical SNR (supernova
A.D. 386) and hence approximately the same age as Kes\sp73 (Vasisht
et al. 1996). Unlike Kes\sp73, it contains a weak, but extended
hard X-ray plerionic core. RCW\sp103 is a virtual twin of Kes\sp73; 
both have radio-quiet, point-like X-ray sources in their centers, share 
morphological similarities, while neither shows evidence for a radio or 
X-ray plerion (see Gotthelf et al. 1997 and refs therein). We infer
a minimum energy in particles and nebular magnetic fields in the Kes\sp73
core to be $E_{min} < 10^{47} \ \rm{d}_7^{17/7}$ erg,
assuming Crab-like parameters and equipartition between magnetic
fields and relativistic particles (Pacholczyk 1970).

The lack of observed plerionic emission can be qualitatively explained
in the following manner. If the Kes\sp73 pulsar has a large dipolar
field $B \sim 10^{15}$ G (VG97), a weak plerion is a natural
consequence at an age of $\sim 2\times 10^3$ yr.  Such a pulsar loses
most of its initial spin energy in a matter of $\sim$ years. The
released pulsar wind would immediately suffer strong adiabatic and
synchrotron losses. This leads to a bright plerion at an early age
($\sim 10 - 100$ yrs) with subsequent rapid decline in surface
brightness.  This effect is shown by Bhattacharya (1990) in his paper
on the morphology of SNR with central pulsars.  He shows that for
pulsars with $B$-fields spanning the range $10^{12}$ G to $1.5 \times
10^{13}$ G, the ones with the highest B have the faintest plerions
at an elapsed time $t\sim 10^3$ yr.  For magnetars, $B \simeq 10^{15}$
G, this rapid decline in surface brightness may be much more
pronounced.

\noindent
{\bf Acknowledgements:} This research has made use of data obtained
through the HEASARC at GSFC. We thank Shri Kulkarni for discussions.  
GV thanks GSFC for hosting him.  EVG's research is supported by NASA.  
GV's research is supported by NASA and NSF grants.

\clearpage

{\noindent Fig. 1. -- Radial profiles of Kes 73 in the hard- and
soft-bands (above and below 2.5 keV). The hard-band profile is
consistent with an unresolved point source, while the soft-band
profile indicates additional diffuse emission. The dotted vertical lines
indicate the regions used in the spectral decomposition (see \S 3).}

{\noindent Fig. 2. -- The SIS spectrum and model of the Kes\sp73
region: The upper panel represents a background subtracted
pulse-height spectrum of the supernova remnant in the entire region
($r \simlt 4'$), and the best model function of a thermal plasma
consisting of a bremsstrahlung continuum including Gaussians emission
lines (see table 2), and a power-law component to represent the compact source
emission. SIS-0 and SIS-1 data have been summed. Crosses denote the
actual data and the histogram indicates the model.  Residuals are shown in the
lower panel with $1\sigma$ error bars (See tables 1 \& 2 for the
fitted parameters).}

\bigskip 

{\noindent Plate 1 -- Images of Kes 73 using data from both SIS
cameras. The brightness distribution is exposure corrected and
smoothed.  The plots are centered on the peak (broad-band) emission
and displayed on a linear scale with contours in 16 uniform
increments, scaled to the image max.  {\bf Top Left} -- SIS image of
Kes 73 in the hard-band (2.5 - 10.0 keV); the image is consistend with
an unresolved SIS point source.  {\bf Top Right} -- SIS image of Kes
73 in the soft-band (0.5 - 2.5 keV); evident is a additional strong
diffuse component.  {\bf Bottom Left} -- The above hard-band image
deconvolved using the Lucy restoration method.  {\bf Bottom Right} --
The above soft-band image deconvolved using the Lucy restoration
method.}

\bigskip

{\noindent Plate 2  -- Decomposed SIS spectra of Kes 73.  ({\bf
Upper panels}) The computed background subtracted spectrum (crosses)
of the central compact source with a single absorbed power law fit
(histogram).  The best fit value for the photon index is $3.4$.
({\bf Bottom panels}) The computed spectrum for the thermal-nebula +
background (crosses), fit with a $kT = 0.6$ keV Raymond-Smith thermal
plasma model (histogram); residuals are apparent around O, Ne, and Mg.
In the energy band below about 2.5 keV, the thermal component dominates, 
whereas the central power-law component is dominant in the harder band. 
The plots are shown scaled identically for ease of comparison.}

\bigskip 

{\noindent Plate 2 (bottom). The Rosat HRI brightness image is
overlayed with the SIS contours of the thermal emission from Kes 73
after subtracting off the non-thermal component.  The method used to
create this image is described in the text. The HRI image is smoothed
with a boxcar filter to bring out features correlated on the same 
spatial scales as the contoured image.}

\clearpage 

\onecolumn

\small

\def\ray{Raymond \& Smith}
\def\elem{(${\rm{^{Mg,Si,S}_{Ar,Ca,Fe}}}$)}

\begin{deluxetable}{lccc}
\tablewidth{250pt}
\tablecaption{Spectral Models and Fit Parameters
\label{tbl-2}}
\tablehead{
\colhead{Model} & \colhead{Continuum} & \colhead{N$_H$} &
\colhead{$\chi^2_{\nu}$ ($\nu$)} \\
 & ($\Gamma$; kT) & \colhead{($10^{22}$ cm$^{-2}$)}   &
}
\startdata
\noalign{\vskip 5pt}
\multispan4{\hfil --- The Compact Source Spectrum --- \hfil}\nl\noalign{\vskip 5pt}
 Power-law         & $3.1 - 3.7$   & $2.7 - 3.4$  & $1.0 (151)$ \nl
 Bremsstralung     & $1.7 - 2.2$   & $1.9 - 2.4$  & $1.0 (151)$ \nl
 Blackbody         & $0.6 - 0.7$   & $1.0 - 1.4$  & $1.0 (149)$ \nl
 ~+ power-law      & $0.9 - 3.6$   &    $ - $     &   $ - $     \nl
\noalign{\vskip 5pt}
 \multispan4{\hfill --- The Source + Nebular Spectrum ---\hfill}\nl
\noalign{\vskip 5pt}
 Bremss + lines    & $0.5 - 0.9$   & $2.1 - 2.2$  & $1.0 (187)$ \nl
 ~+ power-law      & $3.4$ fixed   &     $ - $    &   $ - $     \nl
 R\&S              & $0.6$         &    $ 2.9 $   & $3.2 (207)$ \nl
 ~+ power-law      & $3.4$ fixed   &     $ - $    &   $ - $     \nl
\noalign{\vskip 5pt}
 \multispan4{\hfill --- Decomposed Nebular Spectrum ---\hfill}\nl
\noalign{\vskip 5pt}
 Bremss + lines    & $0.6 - 0.8$   & $1.6 - 2.3$  & $1.0 (74)$ \nl
 R\&S              & $0.57 - 0.62$ & $2.7 - 2.9$  & $3.2 (88)$ \nl
\enddata
\tablenotetext{}{Fits using ASCA SIS data between $0.9-8.0$ keV. All continuum 
fits values are quoted in units of keV; all power-laws slopes 
are given as photon indexes $F_{\nu} \propto \nu^{-\Gamma}$; \ray\ (R\&S)
fits with abundance set to 2 times cosmic. Nebular fits include 11 lines of
Mg, Si, S, Ar, Ca. Gaussian line fits required $\sigma = 20$ eV. Errors are 
the formal 90\% confidence limit for one interesting parameter.}
\end{deluxetable}

\begin{deluxetable}{lcccc}
\tablewidth{250pt}
\tablecaption{Measured Emission Lines in Kes 73 \vfill
\label{tbl-1}}
\tablehead{
\colhead{Line} & \colhead{Lab}    & \colhead{Line}    & \colhead{Norm} \nl
\colhead{Name} & \colhead{Energy} & \colhead{Center}  & \colhead{$10^{-3}$ cts/cm$^{2}$/s}
}
\startdata
 O  He$\alpha$\dotfill    & 0.65         & 0.6$\pm 0.1 $     & $\simlt 0.2$ \nl 
 Ne He$\alpha$\dotfill    & 0.92         & 0.9$\pm 0.1 $     & $\simlt 5$ \nl
 Fe-L blends              &              &                   & $\simlt 0.1$\nl
 Mg He$\alpha$\dotfill    & 1.35         & 1.340$\pm 0.003$  & $4.0  \pm 0.5$\nl
 Mg Ly$\alpha$\dotfill    & 1.48         & 1.47$\pm 0.05$    & $0.14 \pm 0.14$\nl
 Si He$\alpha$\dotfill    & $\sim 1.86$  & 1.848$\pm 0.003$  & $3.1  \pm 0.3$ \nl
 Si Ly$\alpha$\dotfill    & 2.00         & 1.97$\pm 0.02$    & $0.14 \pm 0.06$\nl
 Si He3p+4p\dotfill       & $\sim 2.20$  & 2.20$\pm 0.01$    & $0.27 \pm 0.04$ \nl
 Si Ly$\beta$\dotfill     & 2.38         & 2.37$\pm 0.03$    & $0.25 \pm 0.04$ \nl
 S  He$\alpha$\dotfill    & $\sim 2.45$  & 2.457$\pm 0.005$  & $0.78 \pm 0.06$\nl
 S  Ly$\alpha$\dotfill    & 2.62         & 2.64$\pm 0.14$    & $0.05 \pm 0.03$\nl
 S  He$\beta$\dotfill     & 2.89         & 2.89$\pm 0.03$    & $0.07 \pm 0.03$\nl
 Ar He$\alpha$\dotfill    & $\sim 3.1$   & 3.08$\pm 0.03$    & $0.11 \pm 0.02$\nl
 Ar/Ca He$\alpha$\dotfill & $\sim 3.8$   & 3.8$\pm 5.0$      & $0.01 \pm 0.01$\nl
 \enddata
\tablenotetext{}{A single temperature bremsstrahlung with power-law
model is assumed for the continuum emission. All energies are quoted in keV.
Errors are the formal 90\% confidence limit for one interesting parameter.
Line normalizations are given as total photons per cm$^{2}$ per sec in the line.}
\end{deluxetable}


\onecolumn

\clearpage

\begin{figure}
\centerline{\psfig{file=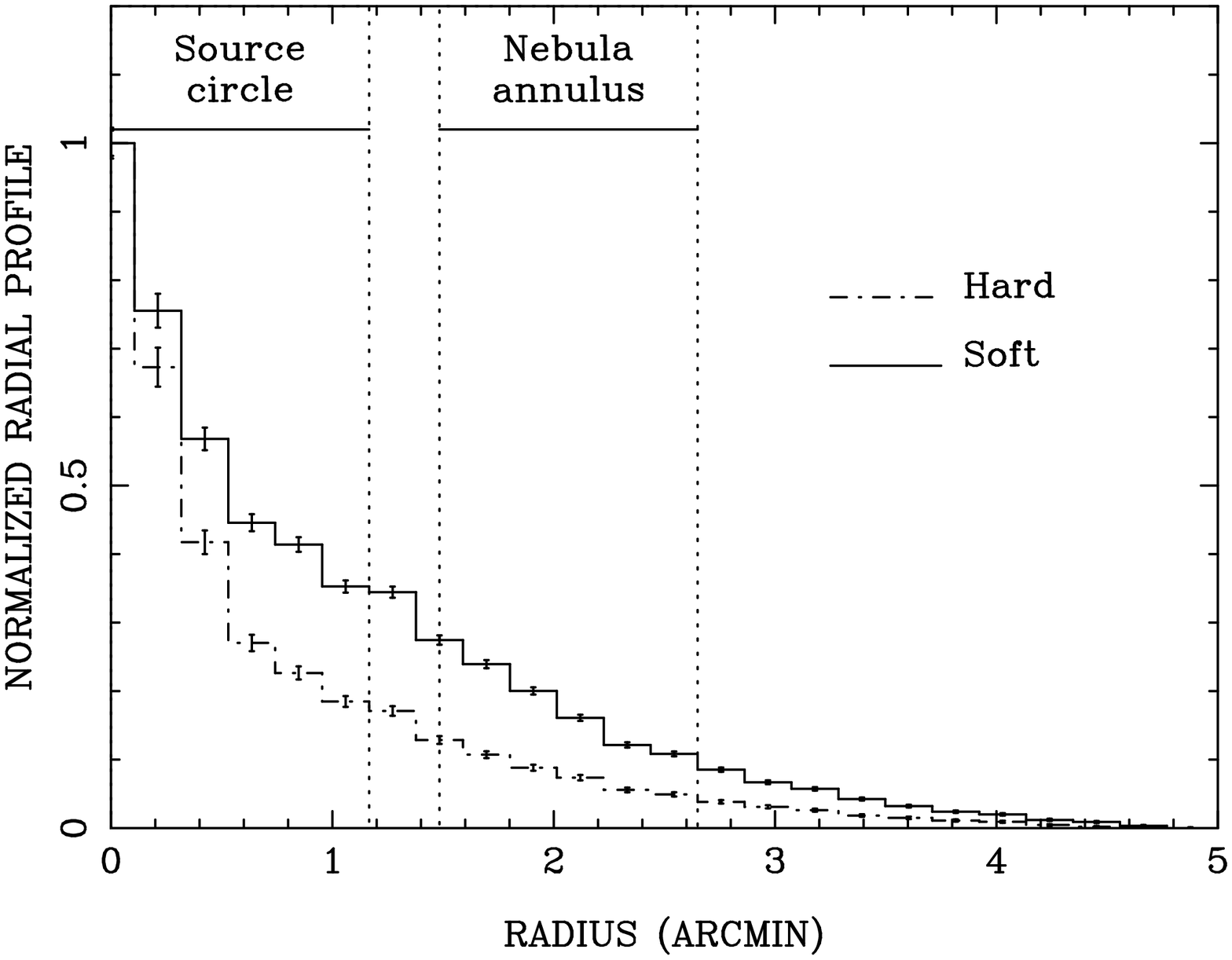,height=4.0in,angle=270.0}}
\bigskip\bigskip\bigskip\bigskip
\centerline{\psfig{file=figure2.ps,width=5.0in,angle=270}}
\end{figure}

\clearpage 

\begin{figure}
\centerline{ \hfill
\psfig{file=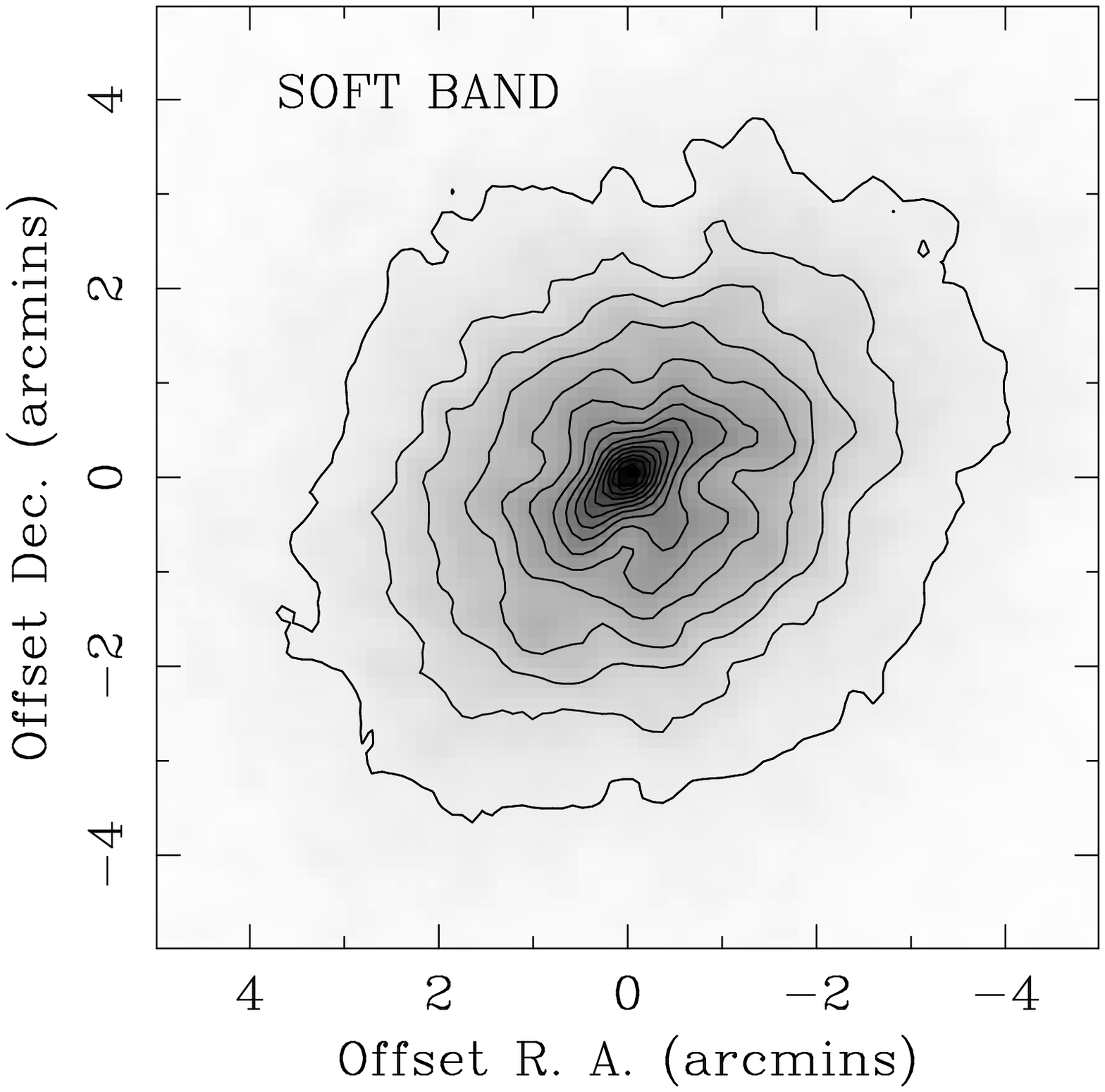,width=3.5in,angle=270.0,bbllx=50bp,bblly=25bp,bburx=487bp,bbury=550bp}
\psfig{file=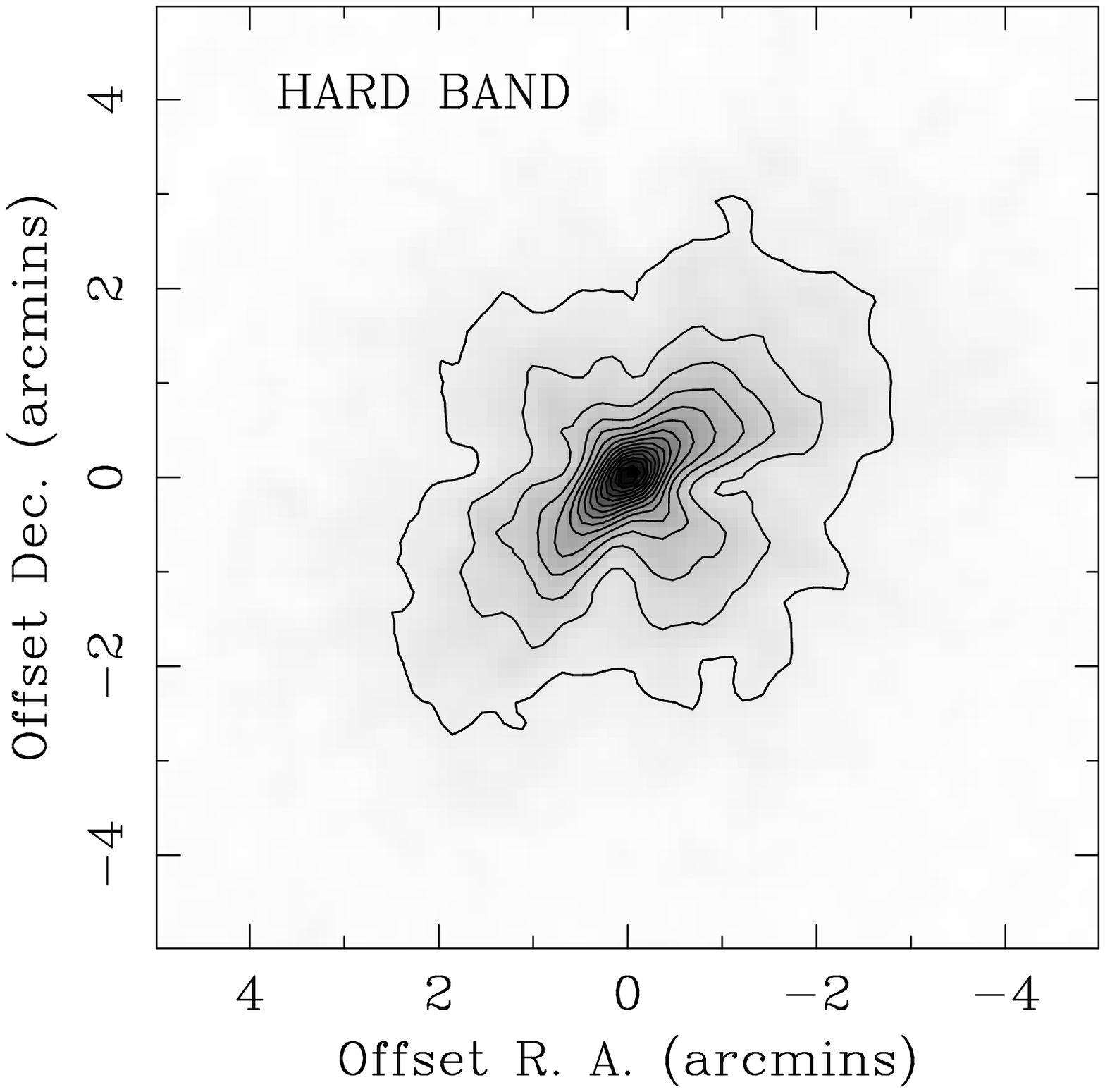,width=3.5in,angle=270.0,bbllx=50bp,bblly=25bp,bburx=487bp,bbury=550bp}\hfill}
\bigskip\bigskip\bigskip\bigskip
\centerline{ \hfill
\psfig{file=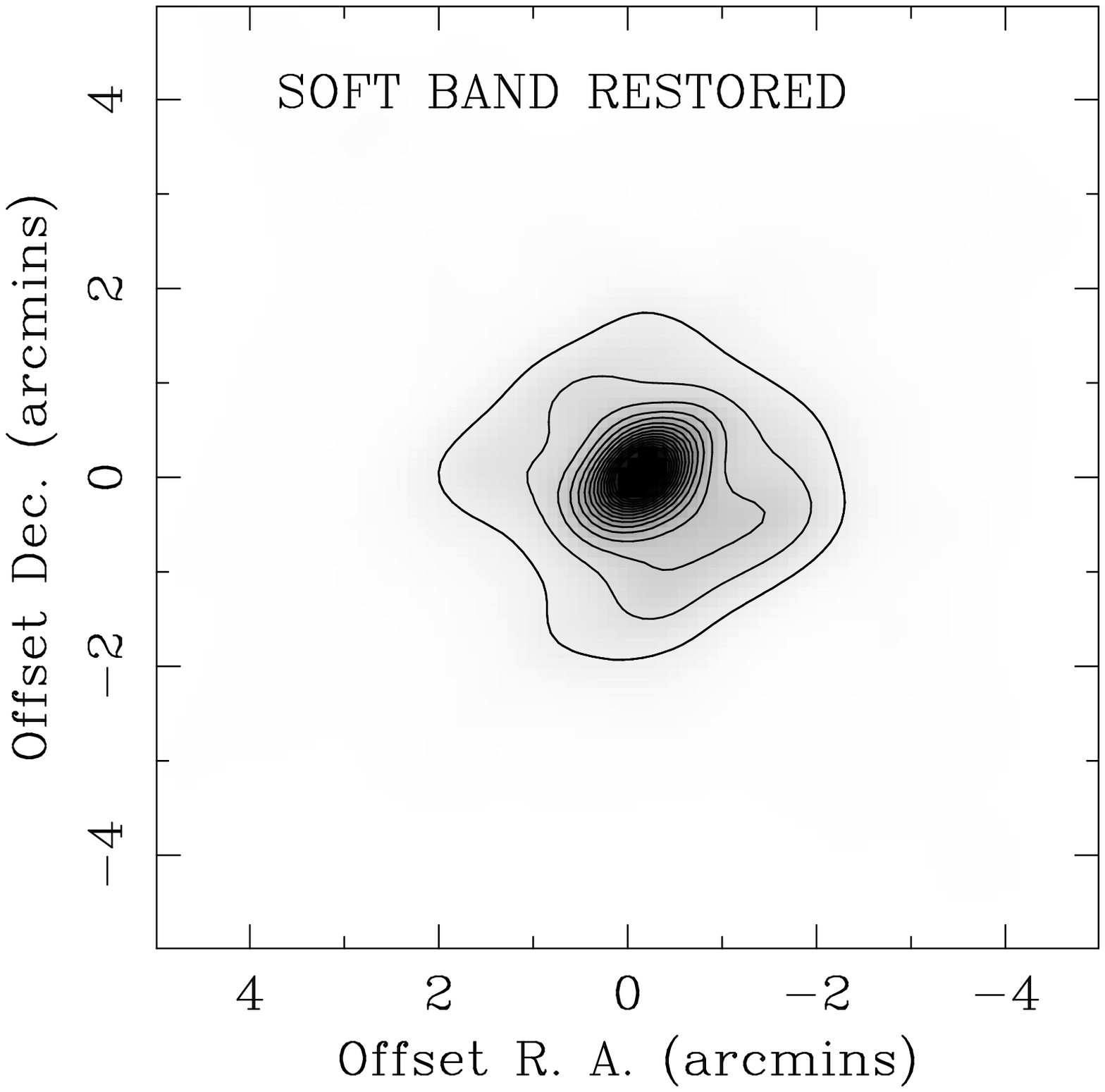,width=3.5in,angle=270.0,bbllx=50bp,bblly=25bp,bburx=487bp,bbury=550bp}
\psfig{file=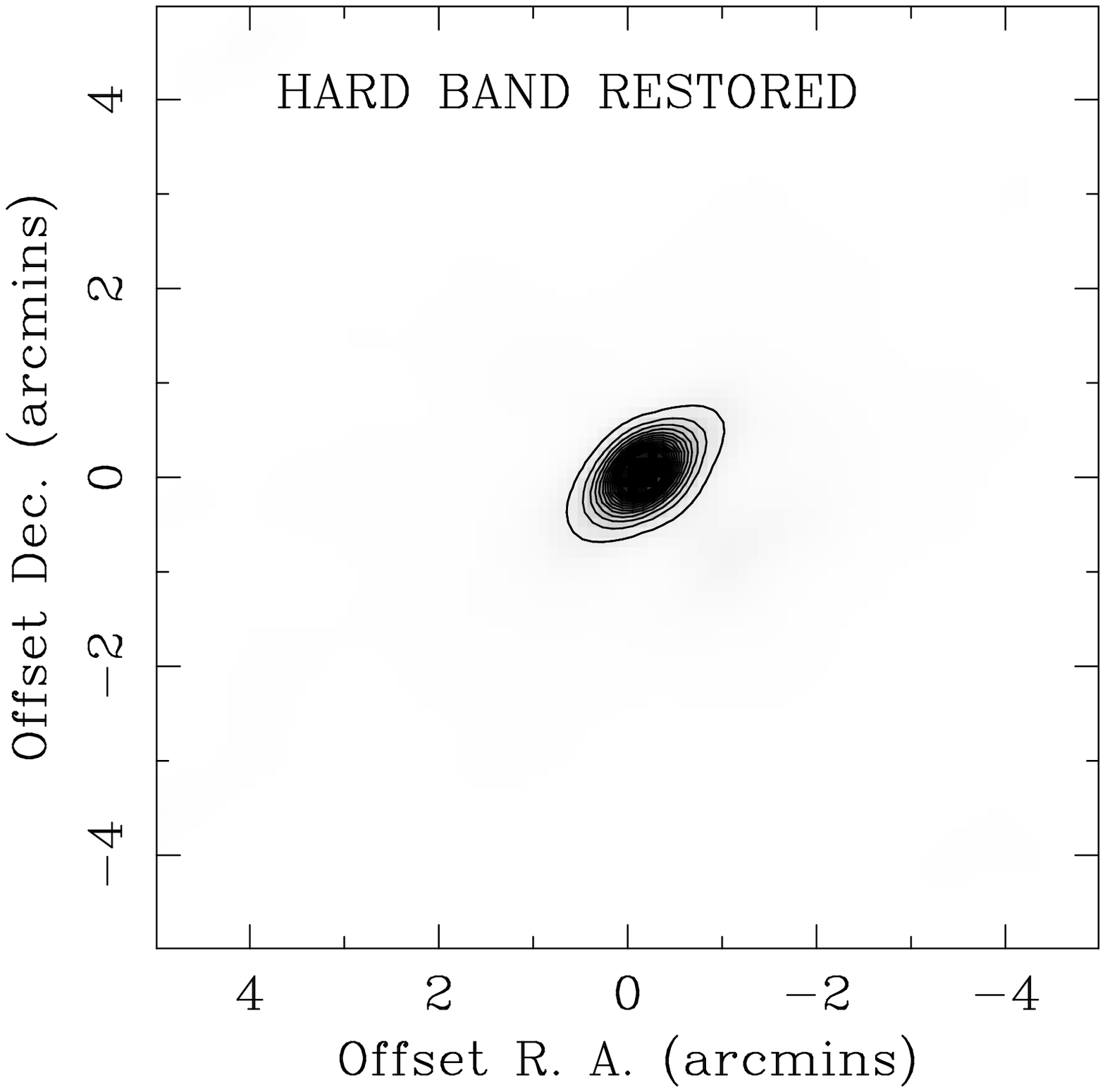,width=3.5in,angle=270.0,bbllx=50bp,bblly=25bp,bburx=487bp,bbury=550bp}\hfill}
\end{figure}

\clearpage 
\begin{figure}
\centerline{
\psfig{file=plate2a.ps,width=4.0in,angle=270.0}
}
\bigskip\bigskip
\centerline{
\psfig{file=plate2b.ps,width=4.0in,angle=270.0}
}
\bigskip\bigskip
\centerline{
\psfig{file=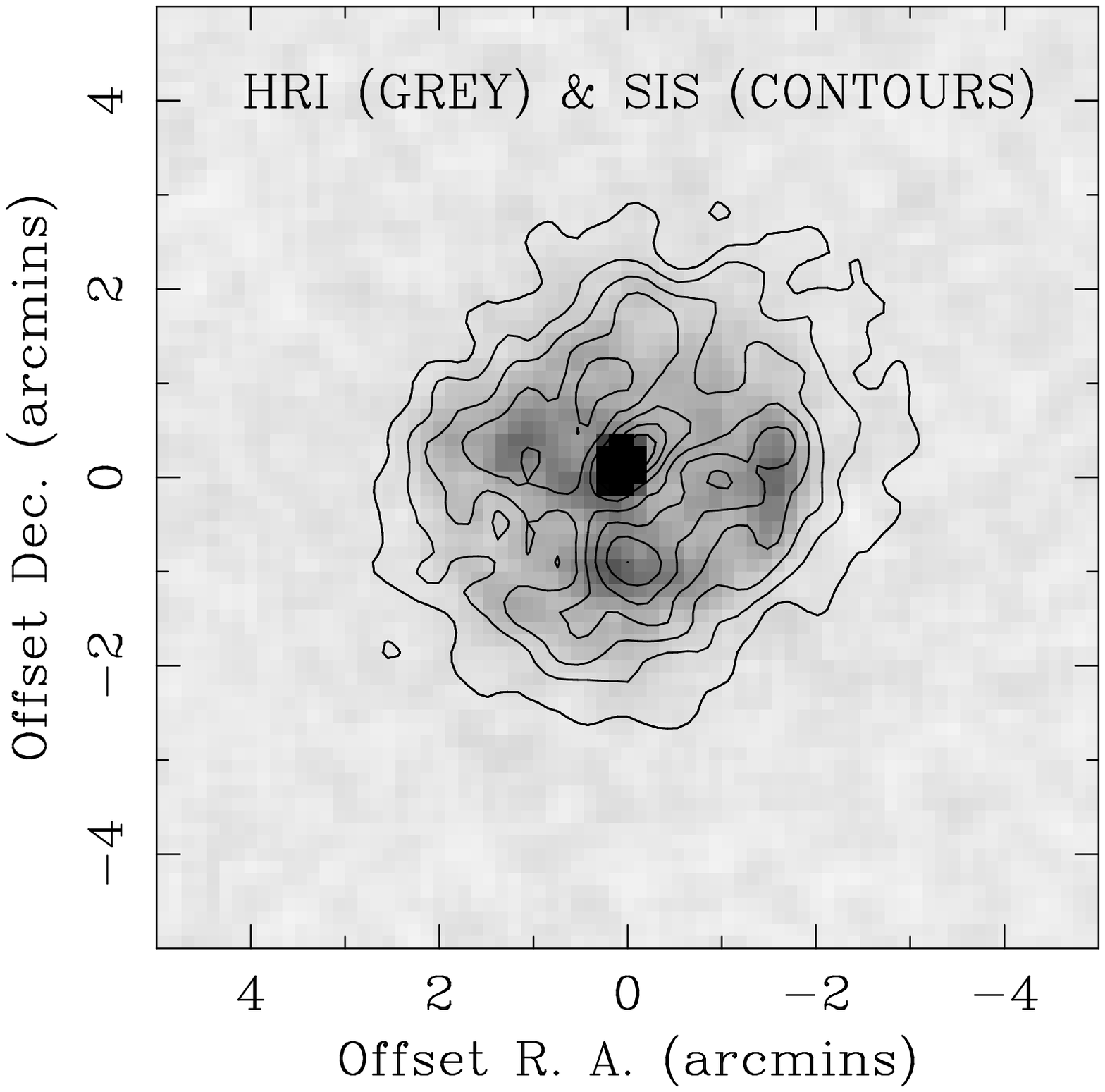,width=3.0in,angle=270.0,bbllx=50bp,bblly=25bp,bburx=487bp,bbury=550bp}
}
\end{figure}

\end{document}